\documentclass[10pt, conference, letterpaper]{IEEEtran}

\usepackage[T1]{fontenc}
\usepackage[latin9]{inputenc}
\usepackage{geometry}
\usepackage{amsmath}
\usepackage{xpatch}
\usepackage{amssymb}
\usepackage{esint}
\usepackage{mathtools}
\usepackage{amsthm}
\usepackage{nicefrac}
\usepackage{bigints}
\usepackage{mathtools}
\usepackage{blkarray, bigstrut}
\usepackage{physics}
\usepackage{calligra}
\usepackage{graphicx}
\usepackage{epstopdf}
\usepackage{dsfont}
\usepackage{array,ragged2e}
\usepackage{enumitem}
\usepackage{etoolbox}
\usepackage{babel}
\usepackage[nopar]{lipsum}
\usepackage{psfrag}
\usepackage[ruled,lined,linesnumbered]{algorithm2e}
\usepackage{algorithmic}
\usepackage{algorithm2e}
\usepackage{balance}
\usepackage{float}
\usepackage{hyperref}
\usepackage{subcaption}

\SetKw{KwBy}{by}

\makeatletter
\newcommand{\removelatexerror}{\let\@latex@error\@gobble}
\makeatother

\graphicspath{ {Figures/} }

\xpatchcmd{\proof}{\hskip\labelsep}{\hskip5\labelsep}{}{}  
\makeatletter
\xpatchcmd{\proof}{\@addpunct{.}}{\@addpunct{:}}{}{}
\makeatother

\renewcommand\[{\begin{equation}}
\renewcommand\]{\end{equation}} 
\usepackage{listings}
\usepackage{fancyvrb}
\usepackage{framed}

\usepackage{courier}
\usepackage[usenames,dvipsnames,table]{xcolor}

\definecolor{dkgreen}{rgb}{0,0.3,0}
\definecolor{gray}{rgb}{0.5,0.5,0.5}

\makeatletter
\newcommand*{\rom}[1]{\expandafter\@slowromancap\romannumeral #1@}
\makeatother

\newlength{\oldtextfloatsep}
\usepackage{siunitx}
\usepackage{tabu}
\usepackage{booktabs}
\usepackage{multirow}
\usepackage{capt-of}
\usepackage{array}
\usepackage{arydshln}
\newcommand{\comment}[1]{}

\usepackage{url}
\begin{document}

\title{
A Multi-Agent Deep Reinforcement Learning Approach for RAN Resource Allocation in O-RAN
}

\author{\IEEEauthorblockN{
Farhad Rezazadeh\IEEEauthorrefmark{1}\IEEEauthorrefmark{2},
Lanfranco Zanzi\IEEEauthorrefmark{3},
Francesco Devoti\IEEEauthorrefmark{3},
Sergio Barrachina-Mu\~noz\IEEEauthorrefmark{1},
Engin Zeydan\IEEEauthorrefmark{1},\\
Xavier Costa-P\'erez\IEEEauthorrefmark{3}\IEEEauthorrefmark{5},
and Josep Mangues-Bafalluy\IEEEauthorrefmark{1}
}
\IEEEauthorrefmark{1}\normalsize{}Centre Tecnol\'ogic de Telecomunicacions de Catalunya (CTTC), Barcelona, Spain\\
\IEEEauthorrefmark{2}Universitat Polit\'ecnica de Catalunya (UPC), Barcelona, Spain\\
\IEEEauthorrefmark{3}NEC Laboratories Europe, Heidelberg, Germany\\
\IEEEauthorrefmark{5}i2CAT Foundation and ICREA, Barcelona, Spain\\
{\normalsize{}Contact Emails:  \texttt{\{name.surname\}@cttc.es}, \texttt{\{name.surname\}@neclab.eu}
}
}

\maketitle

\begin{abstract}

Artificial intelligence (AI) and Machine Learning (ML) are considered as key enablers for realizing the full potential of fifth-generation (5G) and beyond mobile networks, particularly in the context of resource management and orchestration.
In this demonstration, we consider a fully-fledged 5G mobile network and develop a multi-agent deep reinforcement learning (DRL) framework for RAN resource allocation. 
By leveraging local monitoring information generated by a shared gNodeB instance (gNB), each DRL agent aims to optimally allocate radio resources concerning service-specific traffic demands belonging to heterogeneous running services. We perform experiments on the deployed testbed in real-time, showing that DRL-based agents can allocate radio resources fairly while improving the overall efficiency of resource utilization and minimizing the risk of over provisioning. 

\end{abstract}

\section{Introduction}
The highly heterogeneous nature of 5G and beyond makes it increasingly challenging to meet the performance requirements for emerging use cases. 
The strict reliability and latency requirements exacerbate the need for intelligent orchestration solutions able to provide resource and cost-efficient services. In this context, network slicing is a key enabler to underpin such challenging scenarios by deploying logically-isolated virtual network instances. Motivated by open architecture and interfaces for intelligent and agile RAN, O-RAN Alliance~\cite{ORAN_White_paper} provides flexibility, interoperable and open interface for virtualization and network innovation.
Despite this revolutionary approach, it is still not clear how to efficiently support large network slicing scenarios. Therefore, we address this challenge by proposing a hierarchical architecture for network slice resource orchestration. In particular, given the variable spatio-temporal distribution of mobile traffic demands~\cite{AztecInfocom20-demo}, we envision the dynamic setup of a network of local decision agents (DAs) as virtual software instances co-located within the Near-Real Time RAN Intelligent Controller (Near-RT RIC) premises. These instances are capable of accessing local RAN monitoring information and extracting local knowledge without the need for a centralized entity to make decisions based on the aggregated information. 

Our framework, which was initially presented and validated by means of an exhaustive simulation campaign in~\cite{Specialization_TVT}, leverages a distributed learning mechanism and multiple decision agents that collaboratively specialize their decision policies onto real-time traffic demands, aided by a coordinated exchange of information to avoid the occurrence of conflicting situations.
In our approach, we perform resource allocation in a distributed fashion directly at the edge of the network. This brings manyfold advantages including $i$) performing radio resource allocation decisions leveraging the most up-to-date information, $ii$) the decision process does not involve a centralized controller, significantly decreasing the need for control information exchange and reducing the overhead towards the core of the network, and
$iii$) by allowing information exchange among local DAs, we can establish federated learning schemes to further enrich the capabilities of the DAs. Namely, DAs not only rely on local observations to drive the learning process but can exploit the experience from other (statistically different) RAN nodes to speed up the convergence time and better generalize resource allocation decisions.


\begin{figure}[t]
\centering
\includegraphics[width=\columnwidth, clip, trim={0cm 0cm 0cm 0cm}]{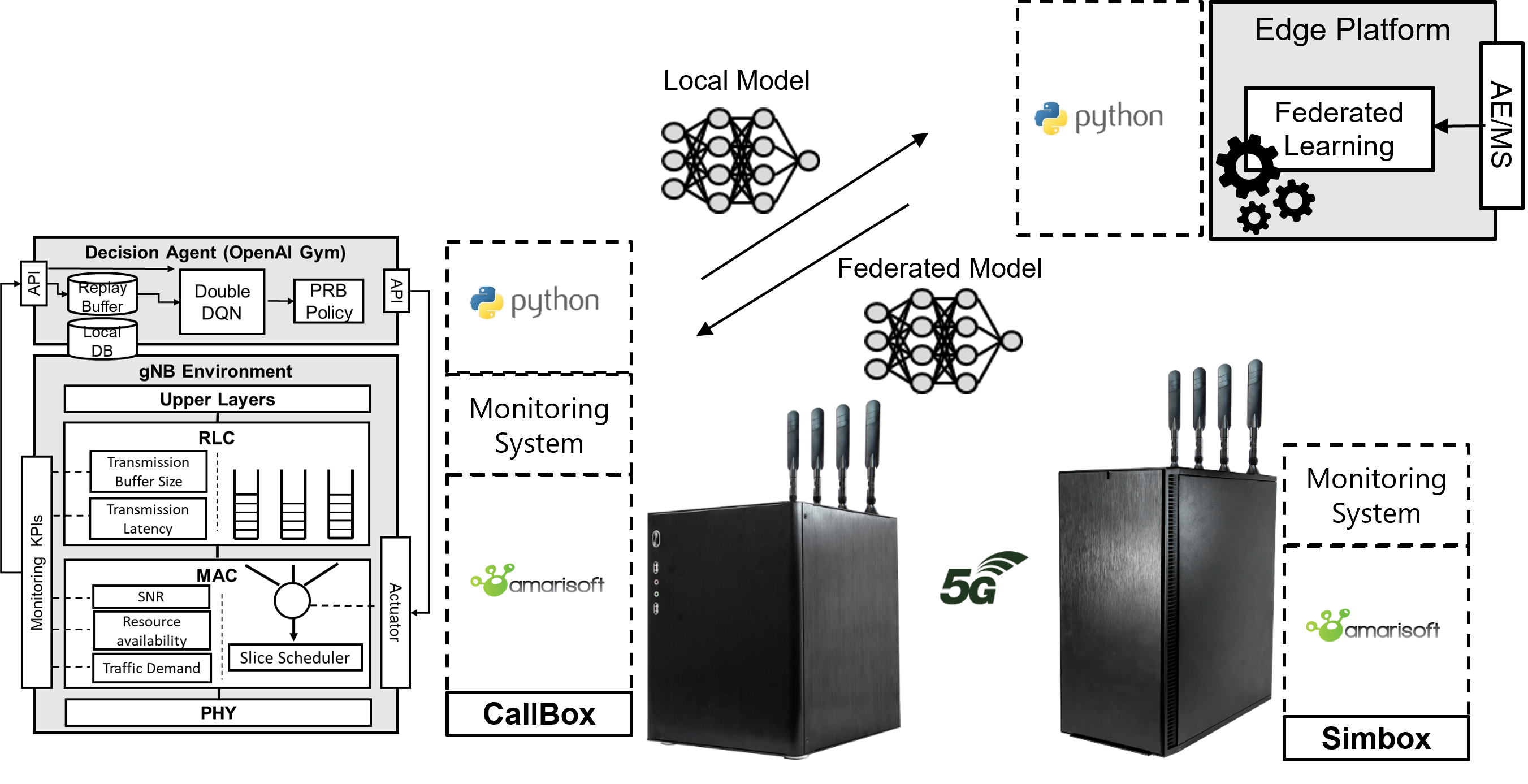}
\caption{Federated RAN resource allocation architecture.}
\label{fig:testbed_architecture}
\end{figure}

\section{Testbed Architecture}
\begin{figure}[t!]
\centering
\includegraphics[width=.5\columnwidth, clip,trim={0cm 0cm 0cm 0cm}]{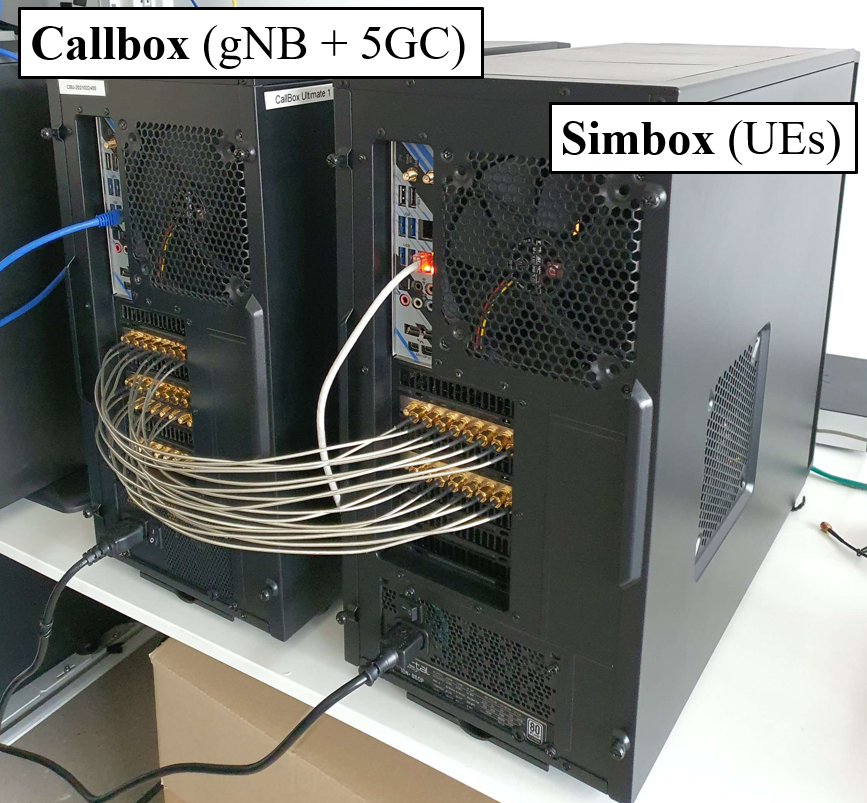}
\caption{Testbed deployment.}
\label{fig:testbed_real}
\end{figure}

We propose a distributed architecture for RAN slice resource orchestration based on DRL, consisting of multiple AI-enabled decision agents that independently take local radio allocation decisions without the need for a centralized control entity.
We design and implement the overall framework as shown in Fig.~\ref{fig:testbed_architecture}. The testbed includes:\\\
\textbf{AMARI UE Simbox}: It is capable of simulating tens of User equipments (UEs) sharing the same spectrum with different types of traffic within multiple cells. Each UE can be independently configured as a 5G NR device \cite{AM_simbox}.\\
\textbf{AMARI Callbox}: provides Enhanced Packet Core (EPC)/5G Core (5GC) functionalities, including the authentication of UEs. Moreover, it implements up to 3 gNBs cells, enabling functional and performance testing. Thanks to its multi-cell configuration, it is also suitable for handover and reselection tests. The technical specifications of the gNB and the core can be found on the vendor's website \cite{AM_callbox}.\\
\textbf{Monitoring System}: Enables real-time monitoring of multiple KPIs, including the number of connected devices, bandwidth utilization, 
etc., which are stored in a local database.\\
\textbf{Local Decision Agents}: collect and consume local monitoring information from to adjust radio resource allocation policies according to real-time traffic variations. Decision-making is supported by AI algorithms and DRL approaches.\\
\textbf{Federated Learning Layer}: It acts as an aggregation point for the local decision engines. It collects locally trained (and therefore heterogeneous) decision models and combines them to gain global knowledge about the underlying infrastructure behavior to improve the generalization of the decision process in the agents. The overall testbed is depicted in Fig.~\ref{fig:testbed_real}.

\section{Operational Phases}
A simplified example of our demonstration is available
online\footnote{\url{https://www.youtube.com/watch?v=Hw4kcaVk94I}}.
This video presents the network performance of different deployed slices from our monitoring dashboard based on Grafana. 
The possibility of acquiring instantaneous and effective resource allocation in network slicing is precluded due to the heterogeneous and dynamic characteristics of traffic requests. This scenario leverages a Multi-DRL mechanism to alleviate this challenge. Each agent deployed over the Callbox is equipped with a double deep Q-network (DDQN)~\cite{Specialization_TVT} that, after an initial exploration and training phase, achieves the right trade-off between optimal allocated radio resources and traffic demand accommodation, i.e., exploitation. To this aim, the agents receive partial observations from the running services or \emph{slices} (e.g., channel quality, consumed resources, etc.) by interacting with each other and with the underlying physical environment.
A reward is calculated as an incentive mechanism based on the actions of all agents, indicating how the agents ought to behave. The designed reward function guarantees adequate performances, impeding over-provisioning and leading to fair resource allocation among the three slices. 

\begin{figure}[t!]
\centering
\includegraphics[width=\columnwidth, clip,trim={0cm 0cm 0cm 0cm}]{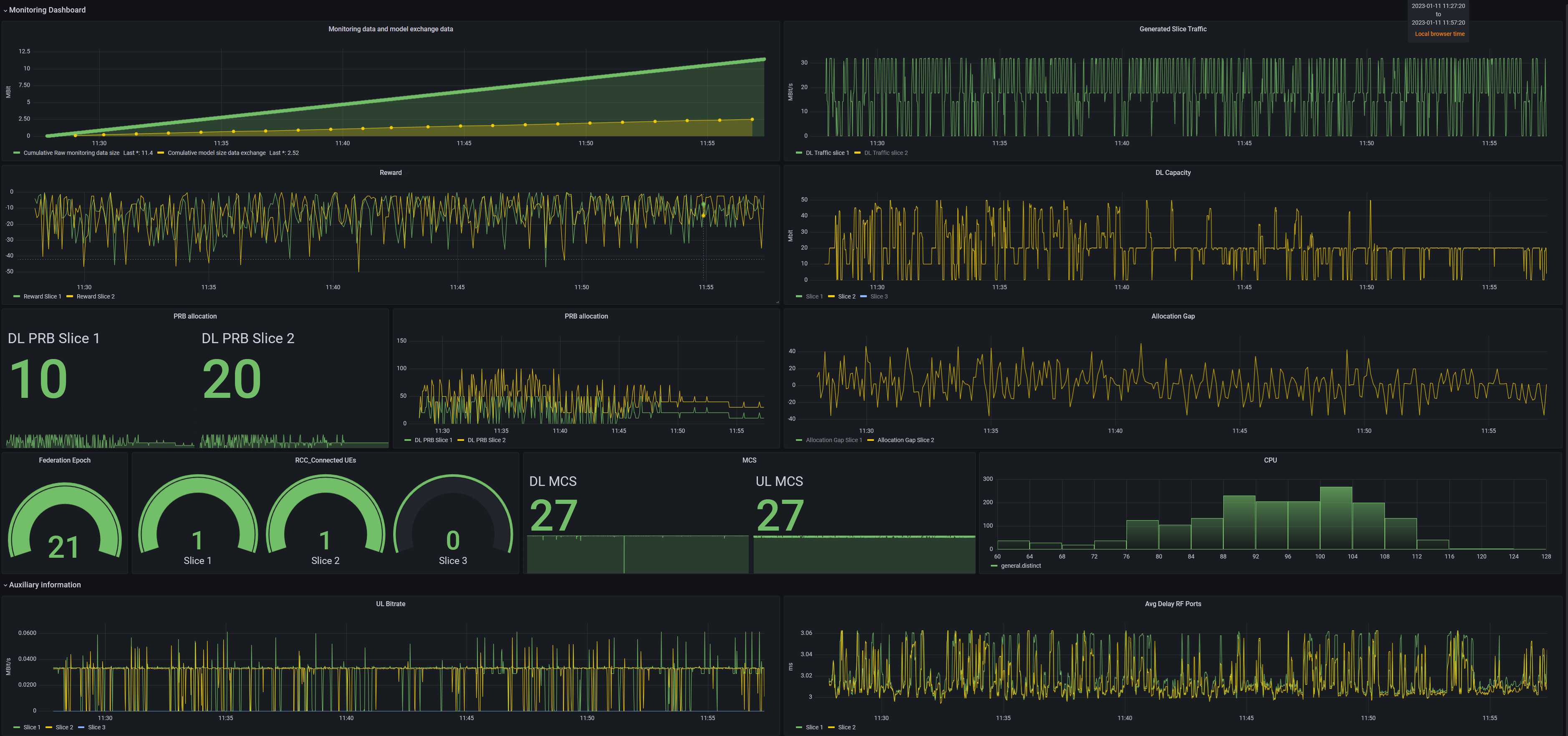}
\caption{Graphical User Interface.}
\label{fig:GUI}
\end{figure}

Fig.~\ref{fig:GUI} illustrates the slice performance of three agents and validates the presented framework by displaying different metrics such as average reward, allocation gap, and allocated radio resources. Furthermore, given the spatio-temporal variation of the traffic demand due to end-user mobility proper of realistic RAN deployments, it is not enough to leverage the geographical locations and related spatial proximity of the gNB to obtain a comprehensive view of traffic demands. Therefore, it is hard for a single AI agent to predict resources for the same slices distributed in different gNB. Federated Learning allows training ML models across multiple decentralized entities, each agent having access to a limited set of overall data realization and statistics.
In this regard, we pursue an experimental approach deploying three enhanced Mobile Broadband (eMBB) slices with different traffic shapes, emulating the slice traffic generated by different gNB, and finally evaluating different federated DRL approaches. 

\section*{Acknowledgment}
The research leading to these results has been partially supported by the H$2020$ MonB5G Project (grant 871780), 
the European Union Smart Networks and Services (SNS) under Horizon-JU-SNS-2022 program (project BeGREEN - grant 101097083), and EdgeDT.
The work is also partly funded by Program UNICO I+D funded by MCIN/AEI/ 10.13039/501100011033 under Grant TSI-063000- 2021-54/55, and in part by the ERDF "A way of making Europe" under Grant PID2021-126431OB-I00, and in part by Generalitat de Catalunya (grant 2021SGR-00770).

\end{document}